\documentclass[%
 reprint,
 amsmath,amssymb,
 aps, nofootinbib
]{revtex4-2}
\usepackage{dcolumn}
\usepackage[hidelinks,citecolor=blue,urlcolor=blue,colorlinks=true,bookmarks=false,hypertexnames=true]{hyperref} 
\usepackage{comment}
\usepackage{amssymb,amsmath,latexsym,graphicx, bm, mathrsfs}
\usepackage[hidelinks,citecolor=blue,urlcolor=blue,colorlinks=true,bookmarks=false,hypertexnames=true]{hyperref} 
\usepackage{natbib} 
\usepackage{xr-hyper}
\usepackage{hyperref}
\setcitestyle{numbers,square} 
\usepackage{graphics,graphicx, cancel}
\usepackage[dvipsnames]{xcolor}
\usepackage{graphicx}
\graphicspath{ {./images/} }
\usepackage[british]{datetime2}
\usepackage[page,toc,titletoc,title]{appendix}
\usepackage{tocloft}
\usepackage{blindtext}
\usepackage{dsfont}
\usepackage{dirtytalk}

\usepackage{comment}
\usepackage{caption}
\usepackage{enumitem}
\usepackage{mathrsfs}
\usepackage{alphalph} 
\usepackage{changepage}
\newcounter{mainthm}
\newtheorem{Proposition}{Proposition}

\newcounter{subthm}[mainthm]

\newtheorem{subtheoreminner}{Theorem}[subthm]

\begin{document}

\title{Stationary Stars Are Axisymmetric in Higher Curvature Gravity}

\author{Nitesh K. Dubey$^{1,2}$}
\email{nitesh.dubey@iiap.res.in}

\author{Sanved Kolekar$^{1,2}$}
\email{sanved.kolekar@iiap.res.in}

\author{Sudipta Sarkar$^{3}$}
\email{sudiptas@iitgn.ac.in}

\affiliation{$^{1}$Indian Institute of Astrophysics, Block 2, 100 Feet Road, Koramangala, Bengaluru 560034, India}

\affiliation{$^{2}$Pondicherry University, R.V. Nagar, Kalapet, Puducherry-605014, India }

\affiliation{$^{3}$Indian Institute of Technology Gandhinagar, Gujarat 382055, India}

\date{\today}

\begin{abstract}
The final equilibrium stage of stellar evolution can result in either a black hole or a compact object such as a white dwarf or neutron star. In general relativity, both stationary black holes and stationary stellar configurations are known to be axisymmetric, and black hole rigidity has been extended to several higher curvature modifications of gravity. In contrast, no comparable result had previously been established for stationary stars beyond general relativity. In this work we extend the stellar axisymmetry theorem to a broad class of diffeomorphism invariant metric theories. Assuming asymptotic flatness and standard smoothness requirements, we show that the Killing symmetry implied by thermodynamic equilibrium inside the star uniquely extends to the exterior region, thereby enforcing rotational invariance. This demonstrates that axisymmetry of stationary stellar configurations is not a feature peculiar to Einstein gravity, but persists even in the presence of higher curvature corrections.
\end{abstract}

\maketitle


\section{\label{sec:level1}Introduction}

Symmetry principles play a defining role in gravitational physics, constraining both the dynamics and the possible equilibrium configurations of self-gravitating systems. Within general relativity, this structure is exemplified by the famous uniqueness theorems, which demonstrate that all asymptotically flat, stationary, and rotating electrovacuum black hole solutions of the Einstein equations must be axisymmetric \cite{Hawking1972, PhysRevLett.26.331, Heusler1998}. These results imply a remarkable rigidity: once the global charges in a black hole spacetime such as mass, angular momentum, and electric charge are fixed, the spacetime geometry is uniquely determined.

A parallel conclusion holds for self-gravitating matter distributions. Assuming a viscous, heat-conducting fluid interior in a stationary, nonsingular, globally hyperbolic, asymptotically Minkowskian spacetime, the stationary stellar configurations are likewise required to be axisymmetric \cite{Lindblom}. The close correspondence between these two classes of results, one describing vacuum spacetimes with horizons and the other extended matter configurations, suggests a deeper universality in the structure of gravitational equilibrium. Axisymmetry appears not as an incidental property but as a manifestation of an underlying geometrical constraint inherent to the dynamics of gravity. Remarkably, in the black hole case, this aspect has also been shown to persist, at least perturbatively, within several higher curvature and modified theories of gravity \cite{Hollands:2022ajj}. Such findings raise a natural but unexplored important question: does an analogous result hold for a stationary stellar equilibrium when Einstein’s theory is generalised? In other words, is the axisymmetry of stationary stars a universal feature of gravitational systems, or a specific consequence of the Einstein field equations?

From a broader theoretical standpoint, higher curvature extensions of general relativity arise naturally in the effective field theory description of gravity. Since the Einstein–Hilbert action is perturbatively nonrenormalizable, quantum corrections would generate an infinite tower of curvature operators suppressed by the ultraviolet scale. These terms are therefore not optional modifications but unavoidable contributions obtained by integrating out short distance degrees of freedom, providing a systematic and diffeomorphism invariant parametrization of deviations from Einstein gravity. In higher dimensional theories, the requirement of second order field equations uniquely selects the Lovelock class \cite{Lovelock:1971yv, Padmanabhan:2013xyr}. They further appear as leading corrections in string theory, where curvature scales approach the quantum gravity regime \cite{Zwiebach:1985uq, Gross:1986mw}. Phenomenologically, such terms offer a controlled arena to test the robustness of classical results, including black hole thermodynamics \cite{PhysRevD.38.2434, Cai2004, Kolekar:2012tq, Ghosh:2020dkk, Sarkar:2019xfd} , and help identify which features of gravitational dynamics are genuinely universal.

Modern approaches to quantum gravity, most notably String Theory and M-theory, naturally require additional spatial dimensions and generically predict higher-curvature corrections to the classical action. Likewise, braneworld constructions such as the Randall–Sundrum model suggest that our observable universe may be embedded in a higher-dimensional bulk, leading to modified effective gravitational dynamics on astrophysical scales. Although still speculative, these frameworks can alter the strong-field behavior of gravity and potentially leave imprints in the structure and phenomenology of compact objects, as well as in gravitational-wave signals detected by facilities like LIGO and the Virgo Collaboration. Establishing (or refuting) the persistence of axisymmetry in such generalized settings is crucial for determining which features of stationary compact objects are truly universal and which depend specifically on the special structure of four-dimensional Einstein gravity.

The present study addresses the fundamental question of the validity of the axisymmetry of equilibrium stellar configurations beyond general relativity, particularly for higher curvature frameworks. Our objective is to determine the precise conditions under which axisymmetry remains an inevitable feature of equilibrium and to clarify the extent to which the geometric rigidity so intrinsic to general relativity continues to hold in more general theories of gravity. We demonstrate that the axisymmetry of stellar equilibrium configurations is not a peculiarity of Einstein’s theory alone but indeed extends to the entire Lovelock class of gravities across arbitrary spacetime dimensions. Furthermore, we establish that the result can also be systematically generalized beyond the Lovelock family to encompass a broad class of higher curvature gravitational theories, thereby suggesting a remarkable universality of geometric rigidity of stationary matter configurations even in extended gravitational theories.

\section{Summary of Previous Work} \label{sec:summaryofprevious}

To prepare the ground, we briefly summarize the axisymmetry theorem for stellar configurations in general relativity~\cite{Lindblom}. The proof crucially depends on Einstein’s field equations, together with the assumptions of a fluid with nonzero coefficients of heat conduction and viscosity in a stationary, nonsingular, globally hyperbolic, asymptotically Minkowskian spacetime. The field equations constrain the geometry strongly enough that any stationary stellar configuration must admit an additional axial Killing vector field. Thus, the dynamics of gravity combined with the matter conditions impose a rigid constraint on the symmetry of equilibrium states. Since the proof explicitly relies on the Einstein equations, we revisit its key steps before extending it to modified gravity theories.

Ref. \cite{Lindblom} begins with two central results concerning relativistic fluids in stationary spacetimes. First, a nonsingular, asymptotically flat, stationary spacetime admitting a Cauchy surface necessarily describes a fluid in thermodynamic equilibrium. Stationarity implies the existence of a timelike Killing vector $\eta^{a}$ along which all physical fields, including the metric $g_{ab}$ and entropy current $s^{a}$, are Lie transported. Defining the total entropy as an integral over spacelike hypersurfaces generated along $\eta^{a}$, Gauss’s theorem and $\nabla_{a}s^{a}\ge0$ together yield $\nabla_{a}s^{a}=0$, establishing thermodynamic equilibrium.

The second result examines the implications of equilibrium for viscous, heat-conducting fluids. Starting with stress-energy tensor for the fluid as
\begin{align}
T_{ab} = \rho u_{a}u_{b} + (p - \zeta \theta)h_{ab} - 2\eta\sigma_{ab} + q_{a}u_{b} + q_{b}u_{a},
\end{align}
vanishing entropy production enforces $\sigma_{ab}=0$, $\theta=0$, and $q_{a}=0$, implying absence of shear, expansion, or heat flow. Here, $\rho$ denotes the energy density, $p$ the pressure, and $\eta$ and $\zeta$ the coefficients of viscosity. By the Eckart theorem~\cite{PhysRev.58.919}, this leads to acceleration along the stationary flow lines $a_{a}\propto\nabla_{a}T$ where $T$ is the Tolman temperature, so that $\xi^{a}\equiv u^{a}/T$ satisfies the Killing equation. Hence, the fluid flow aligns with a spacetime symmetry, and the equation of state becomes barotropic.

The next crucial step is to extend this Killing vector from inside of the configuration to outside. It is this step which requires the use of Einstein's field equations, particularly the vacuum equation $R_{ab} = 0$. Applying $\Box\xi^{a}=0$ and invoking the Cauchy–Kowalewsky theorem~\cite{Folland1995PDE}, Ref. \cite{Lindblom} uniquely extended $\xi^{a}$ beyond the stellar surface, showing via Holmgren’s theorem~\cite{CourantHilbert1962} that it remains a Killing vector in the exterior vacuum. After that, using asymptotic symmetry arguments associated with the Poincaré group as well as analyticity of the spacetime, this vector was identified as the generator of azimuthal symmetry. 

These results continue to hold in the presence of electromagnetic fields under ideal magnetohydrodynamic conditions, where the magnetization is advected with the stellar four-velocity.

\section{Generalisation to Higher-Curvature Theories}

Consider a local diffeomorphism invariant metric theory of gravity whose Lagrangian
density is a scalar constructed from the metric, the Riemann tensor, and its covariant
derivatives \cite{PhysRevD.50.846,Peng:2023yym},
\begin{equation}\label{eq:generalLagran}
L = L\!\,\big(
g^{jk},\,
R_{abcd},\,
\nabla_{e}R_{abcd},\,
\ldots,\,
\nabla_{(e_{1}}\!\cdots\nabla_{e_{m})}R_{abcd}
\big),
\end{equation}
where parentheses denote total symmetrisation with unit weight.  
This symmetrisation is imposed because any antisymmetric combination of covariant
derivatives can be rewritten using the commutator identity for covariant derivatives,
which merely generates additional curvature terms.  

The vacuum field equations for a theory defined by the Lagrangian
\eqref{eq:generalLagran} take the form \cite{Peng:2023yym}
\begin{equation}\label{eq:E-raw}
E_{ab}
= -\tfrac12 g_{ab} L
+ P_{a}{}^{cde} R_{bcde}
- 2\nabla^{c}\nabla^{d} P_{cabd}
+ W_{ab}
= 0.
\end{equation}
Here,
\begin{equation}\label{eq:P-def}
P^{cdef}
\equiv 
\sum_{i=0}^{m}
(-1)^{i}\,
\nabla_{(e_{1}}\!\cdots\nabla_{e_{i})}
Q^{\,e_{1}\cdots e_{i} cdef},
\end{equation}
with
\begin{equation}
Q^{\,e_{1}\cdots e_{i} cdef}
\equiv
\frac{\partial L}{
\partial\! \,\big(
\nabla_{(e_{1}}\!\cdots\nabla_{e_{i})}R_{cdef}
\big)},
\qquad
i = 0,\ldots,m.
\end{equation}
The tensor $W_{ab}$ consists of terms built from covariant derivatives of
$Q^{\,e_{1}\cdots e_{i} cdef}$ together with curvature tensors.
It appears only when the Lagrangian depends explicitly on derivatives of the curvature
$(m \ge 1)$, and its full expression can be found in the Supplementary Materials.
For purely algebraic curvature theories $(m = 0)$, one has $W_{ab} = 0$.

With these preliminaries in hand, we now examine the generalisation of the stellar
axisymmetry theorem and show that the essential result continues to hold in such
higher curvature theories, in close parallel with the situation in Einstein gravity.

\subsection{Existence of the Killing vector}
The first two results discussed in Section \ref{sec:summaryofprevious}, concerning a general-relativistic fluid with nonzero viscosity and heat conduction in thermodynamic equilibrium, remain valid in any diffeomorphism-invariant metric theory of gravity for which the stress--energy tensor is covariantly conserved, \(\nabla_a T^{ab} = 0\), and the Einstein equivalence principle holds. This encompasses Einstein's general relativity, Lovelock gravity, metric \(f(R)\) gravity, and scalar--tensor theories with minimal coupling. More generally, the theorem applies to any minimally coupled diffeomorphism-invariant metric theory of gravity described by a Lagrangian in Eq.\eqref{eq:generalLagran}, if it obeys the Einstein equivalence principle. However, the theorem may fail in theories with nonminimal matter curvature couplings, such as \(f(R,T)\) models, or other modified gravities where \(\nabla_a T^{ab} \neq 0\). Throughout this manuscript, we restrict attention to theories in which this condition holds. We therefore have an additional Killing vector inside the star, which is proportional to the four velocity of fluid particles.\\

In Lovelock theories, the Lovelock action is polynomial in the Riemann tensor, and hence in $\partial^2 g$, so the principal symbol of the field equations depends algebraically on both the background curvature and the coupling constants. So, multiple solution branches may occur, and on certain branches, the principal symbol becomes degenerate \cite{PhysRevD.101.124003, PhysRevD.96.044019, PhysRevD.91.044013}. To understand the analyticity of metric in such a theory, let us consider a globally hyperbolic asymptotically Minkowskian stationary spacetime $(\mathcal{M},g)$ and fix $p\in\mathcal M$. Let us choose stationary coordinates $(x^0,x^\alpha)$ near $p$ adapted to the timelike Killing field $\eta$, so that
\begin{align}
\eta=\partial_{x^0}, 
\qquad 
\partial_{x^0} g_{ab}=0 .
\end{align}
In these coordinates, all $x^0$–derivatives vanish, so the curvature tensor depends only on spatial derivatives of $g$ up to second order. Write $u=(u^A)=(g_{ab})$, viewed as functions of the spatial variables $x^\alpha$ only. On
\begin{align}
\Omega_*:=\{\det(g)\neq 0\}\cap\{g_{00}\neq 0\},
\end{align}
the stationary Lovelock field equations can be written as a second-order partial differential equations system in the spatial variables
\begin{equation}\label{eq:Phi}
\Phi^A\big(x^\alpha,u,\partial u,\partial^2 u\big)=0 .
\end{equation}
Because $g^{-1}$ is rational in $g$ and the Riemann tensor is polynomial in $(g^{-1},\partial g,\partial^2 g)$, each $\Phi^A$ is real-analytic in $(x^\alpha,u,\partial u,\partial^2 u)$ on $\Omega_*$. The dependence on $\partial^2 u$ is generally nonlinear. The above stationary reduction, with an appropriate coordinate choice, is elliptic only on those regions of spacetime (and on those branches of the theory) where the curvature-dependent principal symbol is uniformly invertible and satisfies a quantitative positivity bound; at points of degeneration one expects loss of well-posedness of the stationary boundary value problem, possible change of partial differential equation type, and failure of analytic elliptic regularity. From an effective-field-theory viewpoint \cite{Hollands:2022ajj}, this ``elliptic region'' hypothesis is natural on the Einstein branch in the low-curvature regime where higher-curvature corrections are perturbative, since there the Lovelock principal symbol is a small perturbation of the Einstein principal symbol and uniform ellipticity is stable under such perturbations. We therefore restrict to a nondegenerate (elliptic) branch/region where the stationary, gauge-fixed operator has a uniformly invertible principal symbol with respect to spatial covectors.

Let
\begin{align}
\mathcal{P}^A{}_B(x , \zeta)
=
\frac{\partial \Phi^A}{\partial(\partial_\alpha\partial_\beta u^B)}\,
\zeta_\alpha\zeta_\beta,
\qquad
\zeta\in(\mathbb R^{D-1})^*\setminus\{0\},
\end{align}
denote the principal symbol of \eqref{eq:Phi} in Morrey's sense \cite{Morrey1958Analyticity, Blatt2020}. After imposing a standard local gauge fixing that removes diffeomorphism degeneracy (for example, a DeTurck modification), assume there exists a neighborhood $U$ of $p$ on which $\mathcal{P}(\zeta)(x)$ is invertible for all $\zeta\neq 0$ and satisfies a uniform strong ellipticity bound. Since $g\in C^{3}$, we have $ u\in C^{2+\mu}$, with $0<\mu<1$. The system \eqref{eq:Phi} is analytic and uniformly elliptic on $U$. Hence, by Morrey's analytic regularity theorem for analytic uniformly elliptic second-order systems, $u$ is real-analytic on $U$. Therefore, the stationary vacuum Lovelock metric components $g_{ab}$ are real-analytic in the spatial variables near $p$, and (by stationarity) independent of $x^0$, like the case of Einstein theory discussed in \cite{MuellerZumHagen, MuellerZumHagen2}.

For the general case of the diffeomorphism invariant theories described by the Lagrangian in Eq.\eqref{eq:generalLagran}, in an effective field theory setting, one can assume an expansion of the metric about an analytic stationary Einstein background $g^{(0)}_{ab}$. At each order, the corrections satisfy an elliptic equation with analytic coefficients and an analytic source determined by lower orders. Inductively, each coefficient in the expansion is analytic. While this argument establishes only termwise analyticity of the perturbative solution, it demonstrates that analytic regularity persists provided the background lies in the nondegenerate elliptic stationary class. 

Now, assuming that the stellar surface, say $\Sigma$, is compatible with the exterior geometry and constitutes a non-characteristic surface of the equation
\begin{equation} \label{eq:waveequation}
\Box \xi^a = - R^a{}_b \xi^b,
\end{equation}
the Killing vector can be extended to the exterior region by means of this equation. Since $\xi^a$ satisfies the Killing equation throughout the stellar interior, continuity ensures that Eq.~\eqref{eq:waveequation} also holds on the stellar surface. Using the analyticity of the metric discussed above, we can assume that the coefficients in Eq.~\eqref{eq:waveequation} are analytic on the surface of the star, and we can take $\xi^b$ and $\partial_a \xi^b$ on the stellar surface as analytic initial Cauchy data. The Cauchy--Kovalevskaya theorem then implies the existence of a unique solution just outside the surface of the star \cite{nandakumaran2020partial, CourantHilbert1962, 2008arXiv0806.1036B, ChoquetBruhat}.

\subsection{Properties of the extension beyond the surface }

\subsubsection{Lovelock theory}

We begin with Lovelock theories of gravity. The field equations of a general diffeomorphism invariant metric theory typically involve higher order derivatives and may exhibit various pathologies, including perturbative ghost modes. A well motivated and distinguished subclass is provided by the Lovelock theories, whose equations of motion remain second order and are free from ghost instabilities \cite{Lovelock:1971yv, Zumino:1985dp, PhysRevLett.55.2656}. These theories correspond to Lagrangians that depend only on the metric and the Riemann tensor, that is, the case \(m=0\) in Eq.~\eqref{eq:generalLagran}, together with the defining conditions
\begin{equation}
\nabla_{i} P^{abcd} = 0, \qquad W_{ab} = 0,
\end{equation}
where the index \(i\) may be any index of the tensor \(P^{abcd}\).

To show that the vector field \(\xi\), extended beyond the stellar surface, becomes a Killing vector in the exterior region, we begin with the vacuum field equation of Lovelock theories, given in Eq.~\eqref{eq:E-raw}, which reduces to
\begin{align} \label{lovelockField}
\mathcal{R}^{a}{}_{b}
\equiv
P_{b}{}^{mcd}\, R^{a}{}_{mcd}
= 0 .
\end{align}
The tensor \(P_{b}{}^{mcd}\), defined in Eq.~\eqref{eq:P-def}, results from differentiating the Lovelock Lagrangian with respect to the Riemann tensor. It incorporates the contributions of all Lovelock densities up to the highest non trivial order permitted by the spacetime dimension and shares the algebraic symmetries of the Riemann tensor. Despite being derived from an action that includes higher curvature invariants, the field equations \eqref{lovelockField} remain strictly second order in the metric, thereby avoiding Ostrogradsky instabilities and ensuring a well posed classical theory. The tensor \(\mathcal{R}^{a}{}_{b}\) thus serves as a generalized Ricci tensor, whose vanishing characterizes the vacuum sector of Lovelock gravity.

Since \(\mathcal{L}_{\xi} \mathcal{R}^{a}{}_{b} = 0\), it leads to a differential equation of the deformation tensor  \(t_{ab} \equiv \nabla_{a}\xi_{b} + \nabla_{b}\xi_{a}\). It characterizes how the flow generated by \(\xi_{a}\) locally distorts the metric or nearby worldlines. When \(t_{ab}=0\), the vector field \(\xi_{a}\) is Killing. More generally, \(t_{ab}\) describes the expansion and shear of the associated congruence, with the antisymmetric part removed.

Using the symmetries of the Riemann tensor together with the Lovelock vacuum equations, we obtain the final equation as (see Eq.(I.26) in the Supplementary Materials)
\begin{equation}
\begin{aligned}
0 =\;&
P_{bm}{}^{cd}
\Bigl(
\nabla_{d}\nabla^{a} t_{c}{}^{m}
-
\nabla_{d}\nabla^{m} t^{a}{}_{c}
\Bigr)
\\[3pt]
&\quad
+ \frac12 P_{b}{}^{mcd}
\Bigl(
R^{e}{}_{mdc}\, t^{a}{}_{e}
-
R^{a}{}_{edc}\, t^{e}{}_{m}
\Bigr)
\\[3pt]
&\quad
+ R^{a}{}_{mcd}
\left[
2\,\frac{\partial P_{b}{}^{mcd}}%
{\partial R_{a'm'}{}^{c'd'}}\,
\nabla_{d'}\nabla^{a'} t_{c'}{}^{m'}
-
\frac{\partial P_{b}{}^{mcd}}%
{\partial g^{a'b'}}\,
t^{a'b'}
\right].
\end{aligned}
\label{identitylovelock}
\end{equation}
In the case of general relativity, we have $P_b{}^{mcd} = \left( \delta_b^c\, g^{dm} - \delta_b^d\, g^{cm} \right)/2$
the above identity reduces to the Eq. (21) of \cite{Lindblom}.\\

The linear, second order, coupled system of partial differential equations in
Eq.~\eqref{identitylovelock} for the deformation tensor \( t_{ab} \)
constitutes a Cauchy problem whose initial data are prescribed on a spacelike,
non characteristic hypersurface \( \Sigma \).  As we show below, Holmgren's
uniqueness theorem \cite{CourantHilbert1962, nandakumaran2020partial} implies that the only solution
compatible with vanishing initial data is the trivial one, \( t_{ab}=0 \).\\

The tensor \(t_{ab}\) depends on the vector field \(\xi_{a}\), the metric
\(g_{ab}\), and their first derivatives. We assume the matter configuration is regular in the sense that the 
stress--energy tensor $T_{ab}$ is bounded in the stellar interior 
and vanishes smoothly in the exterior such that the surface $\Sigma$ of the star remains compatible with the exterior geometry. Through the higher-curvature field equations, this means that the spacetime 
$(\mathcal{M}, g_{ab})$ is a non-singular Lorentzian manifold of 
sufficient differentiability for the field equations to be well defined, with all curvature invariants remaining finite throughout the region 
under consideration. At the surface $\Sigma$ of the star we assume the pressure to vanish, while one can allow a bounded jump in the density $\rho$ across the surface. From the junction conditions,
appropriately generalized to Lovelock theories \cite{Papantonopoulos2009, Brassel:2023zle, Guilleminot:2022psn,  Gravanis:2004kx, PhysRevD.62.103502}, then the metric and its
first derivatives are continuous across \(\Sigma\) and all second derivatives are continuous, except for $\partial_n \partial_n g_{ab}$ where $n^i$ is the normal to \(\Sigma\), where there could be a discontinuity. The construction of the
extension ensures that \(\xi^{a}\) and its first derivatives are also
continuous across \(\Sigma\). Consequently, \(t_{ab}\) is continuous across
\(\Sigma\), and since it vanishes in the interior region, it must vanish on
\(\Sigma\) itself.  We next examine the first derivatives of \(t_{ab}\),
\begin{equation} \label{partialderivt}
\partial_{c} t_{ab}
 = \partial_{c}\!\left(
   \xi^{d} \partial_{d} g_{ab}
   + g_{ad}\, \partial_{b} \xi^{d}
   + g_{db}\, \partial_{a} \xi^{d}
 \right).
\end{equation}
In Eq.~\eqref{partialderivt}, the only term containing a second derivative of
the metric is \( \xi^{d} \partial_{c}\partial_{d} g_{ab} \), which is
continuous across \(\Sigma\) since $\xi^a$ is tangent to \(\Sigma\).  To show continuity of the relevant second
derivatives of \(\xi_{a}\), we use Eq.~\eqref{eq:waveequation}.  Expanding its
left-hand side in coordinates gives

\begin{equation} \label{lhsnablaxi}
\begin{aligned}
\nabla_{a}\nabla^{a}\xi^{b}
=\;&
g^{cd}\,\partial_{c}\partial_{d}\xi^{b}
 - R^{b}{}_{c}\,\xi^{c}
\\[2pt]
&\quad
+ \xi^{d}\partial_{d}\!\left(g^{ce}\Gamma^{b}{}_{ce}\right)
 + F^{b}(g,\partial g,\xi,\partial\xi).
\end{aligned}
\end{equation}
where the function \(F^{b}\) depends only on the metric, the vector field,
and their first derivatives, and is therefore continuous across \(\Sigma\).
Combining Eq.~\eqref{eq:waveequation} with Eq.~\eqref{lhsnablaxi} yields an
explicit expression for the second derivatives of \(\xi^{b}\),
\begin{equation}
g^{cd}\partial_{c}\partial_{d}\xi^{b}
 = -\,\xi^{d}\partial_{d}\!\left(g^{ce}\Gamma^{b}{}_{ce}\right)
   - F^{b}(g,\partial g,\xi,\partial\xi),
\end{equation}
whose right-hand side is continuous at \(\Sigma\).  Thus all terms appearing
in the first derivatives of \(t_{ab}\) are continuous across \(\Sigma\), and
\(\partial_{c} t_{ab}\) also vanishes on \(\Sigma\).

The tensors \(t_{ab}\) and \(\partial_{c}t_{ab}\) on \(\Sigma\) therefore
supply Cauchy data for Eq.~\eqref{identitylovelock}.  In a stationary vacuum
spacetime, the metric is analytic in suitable coordinates, and since
Eq.~\eqref{identitylovelock} forms a linear system with analytic coefficients,
Holmgren's uniqueness theorem \cite{CourantHilbert1962, nandakumaran2020partial} guarantees that the
vanishing initial data lead to the unique analytic solution \( t_{ab}=0 \).
Hence, the extension defined by Eq.~\eqref{eq:waveequation} must be a Killing
vector field in the exterior region.

Stationarity implies the existence of an additional Killing vector field
\(\eta^{a}\).  Denoting by \(l^{b}\) the Lie derivative of \(\xi^{b}\) along
\(\eta^{a}\), and using the fact that \(\eta^{a}\) commutes with the wave
operator, we obtain
\begin{align} \label{commutatoriden}
\Box l^{b}
 = \mathcal{L}_{\eta}(\Box\xi^{b})
 = - R^{b}{}_{c}\, l^{c}.
\end{align}
In the last step we have used Eq.~\eqref{eq:waveequation} and the fact that the
Lie derivative of the Ricci tensor vanishes along a Killing vector field.  The
equation admits \(l^{b}=0\) as a solution, and by the uniqueness result
established above, with vanishing Cauchy data on the stellar surface, this
solution is the only one.  Therefore, the two Killing vector fields commute.

\subsubsection{Gravity beyond second order}

The situation becomes more involved once one goes beyond Lovelock's theories. In general, in higher-curvature or effective models, the field equations acquire additional dynamical features. Setting $\mathcal{L}_\xi E_{ab}=0$ in Eq.\eqref{eq:E-raw} gives the linear differential equation with at most $(m+4)$ derivatives of $t_{ab}$ , with $t_{ab} = 0$ as a solution (see Eq.(I.42) of the Supplementary Materials).  In the generic situation---absent Lovelock-type cancellations---this bound is saturated, so the principal part of $\mathcal{L}_\xi E_{ab}$ has differential order $m+4$ acting on $t_{ab}$.  

To prove the uniqueness of the solution $t_{ab} = 0$, we proceed as in the preceding subsection. We assume that the star has a bounded stress–energy tensor with at-most a finite jump across the surface $\Sigma$, one can then have $g \in C^{m+3}$, and no components with differential order $m+4$ that appear are discontinuous except perhaps for the $n^i n^k \partial_{n_i} \partial_{n_k} g_{ab}$, similar to the case in Lovelock theories. Then $\nabla^{r+1}\xi$, and hence $\nabla^{r} t$, are continuous across the stellar surface $\Sigma$ for all $r \le m+3$ (see the supplementary material). Since $t_{ab}=0$ inside the star, continuity gives
\begin{align}
\nabla^r t_{ab}\big|_{\Sigma}=0 \qquad (\forall \; r\le m+3).
\end{align}
Equivalently, one may view $\mathcal{L}_\xi E_{ab}$ as a linear differential operator in $t_{ab}=\mathcal{L}_\xi g_{ab}$ by replacing $\mathcal{L}_\xi g\mapsto t$ and $\mathcal{L}_\xi\Gamma\mapsto \tfrac12\,g^{-1}\!\ast\nabla t$, yielding a homogeneous equation $\mathsf{G}_{ab}[t]=0$ with principal order $m{+}4$. On the spacelike, non-characteristic hypersurface $\Sigma$ then the normal derivatives vanish up to order $(m{+}3)$,
\begin{equation}
(n\!\cdot\!\nabla)^k t_{ab}\big|_{\Sigma}=0 \qquad (k=0,1,\ldots,m{+}3),
\end{equation}
and the intrinsic constraints induced on $\Sigma$ by $\mathsf{G}_{ab}[t]=0$ hold. This bootstrap provides the full vanishing jet $\nabla^r t|_\Sigma=0$ for all $r\le m{+}3$, ensuring the sufficient initial data for the Cauchy problem of order-$(m{+}4)$ $t$–equations (see the discussion below Eq.(I.42) of the Supplementary Materials).

The uniqueness theorem of Holmgren, applied to higher order differential equations, again yields $t_{ab}=0$ as a unique solution just outside the surface of the star, as $t=0$ is a solution of system of PDEs \cite{CourantHilbert1962}. Therefore, $\xi^a$ is a Killing vector field. Furthermore, it commutes with the timelike Killing vector field $\eta^a$, the proof of which is identical to the case of Lovelock theory discussed in the previous subsection. Furthermore, Eq.\eqref{commutatoriden} holds, and $l^b$ is a unique solution for the general theory. So, the Killing vector fields $\xi^b$ and $\eta^b$ commute with each other.

\subsection{Axisymmetry of the exterior spacetime }
In the preceding subsections, we extended the Killing vector field $\xi^a$ into a small open neighborhood in the exterior of the star, in a regime where the metric, as well as the exterior geometry, is analytic. The analyticity of the exterior geometry tells us that the components of $\xi^a$ are also analytic. Therefore, by analytic continuation, the Killing vector fields discussed in the preceding subsections can be extended to cover the full spacetime outside the star. Consequently, the manifold admits two globally defined Killing vector fields, which commute with each other. We now ask: \emph{what symmetry does the isometry generated by these Killing vector fields encode?} To answer this, we assume the spacetime to be asymptotically Minkowski. The Minkowski spacetime possesses the full Poincaré group of isometries; any Killing vector in the physical spacetime must, in that asymptotic region, resemble one of the Poincaré generators.

A spacetime containing a localized matter distribution, such as a star, cannot admit spatial translation or Lorentz boost invariance globally, as these would shift or deform the matter distribution. Invariance under a continuous symmetry requires the existence of a Killing vector field that leaves both the metric $g_{\mu\nu}$ and the matter fields invariant. For a localized star, the stress-energy tensor $T_{\mu\nu}$ is nonvanishing only within a finite spatial region and decays outside the surface, thereby selecting a preferred center and breaking spatial homogeneity. A spatial translation would displace the support of $T_{\mu\nu}$, producing a physically distinct configuration, and thus cannot correspond to an isometry of the spacetime. Similarly, Lorentz boost invariance would require the absence of a preferred rest frame; however, a star defines a natural frame. Under a boost, the matter configuration acquires momentum density and Lorentz contraction, altering $T_{\mu\nu}$ and therefore the associated geometry. Consequently, while such a spacetime may admit time-translation symmetry (if stationary) and rotational symmetry (if axisymmetric), it cannot possess global spatial translation or boost symmetries, reflecting the explicit breaking of Poincaré invariance by localized gravitating matter. Thus, the only reasonable symmetries are those of time translations and spatial rotations. The globally timelike Killing vector field $\eta^a$ accounts for the time-translation symmetry. Consequently, the other Killing vector field $\xi^a$ must represent a linear combination of a rotational generator and the timelike generator. Hence, the spacetime admits rotational invariance about an axis; in other words, it is axisymmetric.

\section{Summary and Conclusion}

In this work, we have shown that the axisymmetry of stationary rotating stars is not a special feature of Einstein gravity but a robust property of a wide class of diffeomorphism invariant metric theories, including Lovelock gravity and higher curvature extensions. The key ingredients behind this result are the conditions of thermodynamic equilibrium, conservation of the stress–energy tensor, and the regularity of the stellar surface. These ensure the existence of a Killing vector aligned with the fluid flow inside the star and allow its unique extension into the exterior vacuum \footnote{The theorem on axisymmetry will also hold for a non-vacuum exterior if the matter distribution outside has sufficient decay property, and such that $\mathcal{L}_\xi T_{ab}=0$ in the exterior.}. We justified that, for Lovelock theories, the metric of a stationary vacuum spacetime is analytic, provided appropriate coordinates and assumptions are adopted, via the Morrey theorem. It then follows that the extension into the exterior vacuum region is unique, as a consequence of the second order nature of the field equations together with standard results on analytic uniqueness. For higher derivative theories, the deformation equations become higher order but remain sufficiently well behaved to guarantee that the extended vector continues to satisfy the Killing equation in the exterior. \\

The combination of this extended Killing vector with the stationary Killing field leads to two commuting symmetries. Asymptotic flatness then plays a decisive role: it restricts the asymptotic symmetry algebra to that of the Poincaré group, leaving only time translations and spatial rotations as compatible with a localized matter distribution. Hence, the second Killing vector must generate a rotational symmetry at infinity, enforcing axisymmetry of the full spacetime.

In precise mathematical terms, we can write down the following proposition:
\begin{Proposition}
Let $(\mathcal{M},g)$ denote a globally hyperbolic asymptotically Minkowskian stationary (nonstatic) spacetime having a viscous heat-conducting fluid confined to a finite region of $\mathcal{M}$. We assume that the coefficients of viscosity and the heat conduction coefficients are positive, and the pressure vanishes on the surface $\Sigma$ separating the exterior vacuum and the confined fluid. Furthermore, also assume that $\Sigma $ is compatible with the exterior geometry. Then, in a diffeomorphism-invariant metric theory of gravity for which the stress--energy tensor is covariantly conserved, \(\nabla_a T^{ab} = 0\), and the Einstein equivalence principle holds, if there exist an analytic atlas with respect to which the metric $g_{ab}$ is an analytic tensor field in a stationary vacuum spacetime then $(\mathcal{M},g)$ will be axisymmetric.      
\end{Proposition}

We allow the spacetime dimension $D \geq 4$. The analysis therefore applies to Lovelock theories in general dimensions, where higher--curvature terms contribute nontrivially to the field equations. 
In four spacetime dimensions, however, the Gauss--Bonnet term is purely 
topological and higher-order Lovelock densities vanish identically, so the 
Lovelock theory reduces to ordinary general relativity. Consequently, in 
$D=4$ our results reproduce the general-relativistic case, while for $D>4$ 
they extend the axisymmetry theorem to genuinely higher-curvature 
gravitational dynamics.

The universal axisymmetry property leads to important observational consequences. In any theory consistent with our assumptions, a rotating star cannot remain stationary while carrying non axisymmetric multipolar structure. A robust detection of persistent violation of axisymmetric would therefore signal new gravitational physics, suggesting either a breakdown of metric theories or violations of the equivalence principle. Such effects could manifest in gravitational wave observations, where such asymmetries would distort the quasinormal mode spectrum, or in high resolution Event Horizon Telescope images, where shadow or emission features could indicate non standard exterior symmetries. In addition, precise tracking of stellar orbits near the galactic centre black hole provides a further channel to probe possible departures from axisymmetry. Thus, confirming axisymmetry of stellar exterior along with the black hole results reinforce confidence across a wide class of effective theories, while its violation would open a rare observational window into non metric gravitational dynamics.
\\

Finally, our results highlight natural directions for further work. The role of boundary conditions deserves closer examination in spacetimes that are not asymptotically flat. In de Sitter space, the absence of spatial infinity and the presence of cosmological horizons complicate the identification of rotational generators. In anti de Sitter space, the richer asymptotic symmetry group and the need for boundary conditions at the conformal boundary may allow or forbid different symmetry extensions. Extending the analysis to these cases, and to more general matter models including strong magnetic fields or dissipative effects, would help clarify how far the universality of axisymmetry extends beyond the scenarios considered here.

\begin{acknowledgments}
This research was initiated while participating in the  International Centre for Theoretical Sciences (ICTS) program - Beyond the Horizon: Testing the black hole paradigm (code: ICTS/BTH2025/03). We also thank Rohit Kumar Mishra for discussion about Holmgren’s uniqueness theorem. The research of S.S. is funded by the Department of Science and Technology, Government of India, through the SERB CRG Grant (No. CRG/2023/000545).

\end{acknowledgments}

\bibliography{apssamp}

\end{document}


\title{Supplementary Materials}

\maketitle

\section{Field equation for the general theory}
Let us consider a local diffeomorphism-invariant metric theory of gravity described by a Lagrangian density that is built from the metric, the Riemann tensor, and its covariant derivatives \cite{PhysRevD.50.846, Peng:2023yym},
\begin{equation} \label{eq:generalLagran}
L =L\!\big(g^{jk},\,R_{abcd},\,\nabla_e R_{abcd},\,\ldots,
\nabla_{(e_1} \cdots \nabla_{e_m)}R_{abcd}\big),
\end{equation}
where parentheses on a string of indices denote total symmetrisation with weight one. The symmetrisation is imposed because antisymmetric combinations of derivatives can be rewritten using commutators of covariant derivatives, which generate additional curvature tensors, so keeping only the symmetrised parts avoids redundancy. We also employ a multi-index shorthand:
\begin{align}
F_i \equiv (e_1\cdots e_i),\qquad 
\nabla_{(F_i)} \equiv \nabla_{(e_1}\cdots\nabla_{e_i)},
\end{align}
and define 
\begin{equation}
Q^{\,e_1\cdots e_i cdef}
\;\equiv\;
\frac{\partial L}{\partial\!\big(\nabla_{(e_1}\cdots\nabla_{e_i)}R_{cdef}\big)},
\qquad i=0,\dots,m,
\end{equation}
so that \(Q^{cdef}\) for \(i=0\) is \(\partial L/\partial R_{cdef}\).

Varying the action
\begin{equation}
S=\int d^D x\,\sqrt{-g}\,L
\end{equation}
with respect to the metric produces three kinds of contributions: the variation of the volume element, the algebraic dependence of \(L\) on \(g^{ab}\), and the dependence through the Riemann tensor and its covariant derivatives. 
The contribution to \(\delta L\) coming from the curvature-derivative dependence is
\begin{equation}
\delta L\Big|_{R,\,\nabla R,\dots}
=\sum_{i=0}^m
Q^{\,e_1\cdots e_i cdef}\;
\delta\big(\nabla_{(e_1}\cdots\nabla_{e_i)}R_{cdef}\big).
\end{equation}
For each fixed \(i\ge1\) we perform integration by parts repeatedly to move the \(i\) symmetrized covariant derivatives off the variation \(\delta R_{cdef}\) and onto the coefficient \footnote{ For instance, in order to separate out a term proportional to $\delta R_{\mu\nu\rho\sigma}$ from $Q^{\lambda\mu\nu\rho\sigma} \, \delta \nabla_\lambda R_{\mu\nu\rho\sigma}$, one is able to achieve this by expressing such a term as
\begin{align*}
Q^{\lambda\mu\nu\rho\sigma} \, \delta \nabla_\lambda R_{\mu\nu\rho\sigma} 
&= Q^{\lambda\mu\nu\rho\sigma} \nabla_\lambda \delta R_{\mu\nu\rho\sigma} 
   - 4 Q^{\alpha\beta\nu\rho\sigma} R_{\lambda\nu\rho\sigma} \, \delta \Gamma^\lambda_{\alpha\beta} \\
&= \nabla_\mu \big(Q^{\mu\alpha\beta\rho\sigma}\, \delta R_{\alpha\beta\rho\sigma}\big)
   - (\nabla_\lambda Q^{\lambda\mu\nu\rho\sigma}) \, \delta R_{\mu\nu\rho\sigma}
   - 4 Q^{\alpha\beta\nu\rho\sigma} R_{\lambda\nu\rho\sigma} \, \delta \Gamma^\lambda_{\alpha\beta}.
\end{align*}
where the following has been used in the first line: \[
\delta(\nabla_\lambda R_{\mu\nu\rho\sigma})
= \nabla_\lambda (\delta R_{\mu\nu\rho\sigma})
- \delta\Gamma^\alpha_{\lambda\mu} R_{\alpha\nu\rho\sigma}
- \delta\Gamma^\alpha_{\lambda\nu} R_{\mu\alpha\rho\sigma}
- \delta\Gamma^\alpha_{\lambda\rho} R_{\mu\nu\alpha\sigma}
- \delta\Gamma^\alpha_{\lambda\sigma} R_{\mu\nu\rho\alpha}.
\]
}. Schematically,
\begin{multline}\label{eq:ibp-general}
\int \! d^Dx\sqrt{-g}\;Q^{\,e_1\cdots e_i cdef}\;
\delta\!\big(\nabla_{(e_1}\cdots\nabla_{e_i)}R_{cdef}\big)
=\int \! d^Dx\sqrt{-g}\;(-1)^i
\nabla_{(e_1}\cdots\nabla_{e_i)}Q^{\,e_1\cdots e_i cdef}\;
\delta R_{cdef}
\\
+\;\text{(boundary terms)} + \text{(commutator curvature terms)}.
\end{multline}
The overall factor $(-1)^i$ arises from moving $i$ fully symmetrized covariant derivatives 
onto $Q$%
\footnote{Each integration by parts on a covariant derivative produces a minus sign,
\[
\int d^Dx\,\sqrt{-g}\, A^a \nabla_a B
= - \int d^Dx\,\sqrt{-g}\, (\nabla_a A^a)\, B \;+\;\text{(boundary)}.
\]
Applying this identity repeatedly for $i$ symmetrised derivatives yields 
the overall factor $(-1)^i$.}.
When commuting covariant derivatives in \eqref{eq:ibp-general}, additional
curvature contractions are generated. Summing the \(i\)-terms (including \(i=0\)) the first term in the right hand side of Eq.\eqref{eq:ibp-general} gives:
\begin{equation}\label{eq:P-def}
P^{cdef}\equiv \sum_{i=0}^m (-1)^i\nabla_{(e_1}\cdots\nabla_{e_i)} Q^{\,e_1\cdots e_i cdef}.
\end{equation}
The tensor $P^{cdef}$ inherits the algebraic symmetries of the Riemann tensor. For the variation of the Riemann tensor, we recall the compact identity
\begin{equation}\label{eq:deltaR-four-short}
\delta R_{abcd}
=\tfrac12\big(
\nabla_c\nabla_b \,\delta g_{ad}
+\nabla_d\nabla_a \,\delta g_{bc}
-\nabla_c\nabla_a \,\delta g_{bd}
-\nabla_d\nabla_b \,\delta g_{ac}\big)
+\delta g_{e[c} R^{e}{}_{d]ab}
+\delta g_{e[a} R^{e}{}_{b]cd},
\end{equation}
which already incorporates the commutator identities introduced above. Substituting \eqref{eq:deltaR-four-short} into $\int P^{abcd}\delta R_{abcd}$, integrating by parts, and using the symmetries of $P^{abcd}$, one obtains the standard result
\begin{equation}\label{eq:IBP-result}
\int d^Dx\sqrt{-g}\;P^{abcd}\,\delta R_{abcd}
=\int d^Dx\sqrt{-g}\Big(
-2\,\nabla^{c}\nabla^{d}P_{cabd}
+P_{a}{}^{cde}R_{bcde}
\Big)\delta g^{ab}
+\;\text{(b.t.)}.
\end{equation}

\medskip

The remaining terms can be combined to give a tensor $W_{ab}$ that appear only when the Lagrangian depends on covariant derivatives of the curvature ($m \ge 1$). For $m = 0$, we have $W_{ab} = 0$. The tensor $W_{ab}$ admits a finite decomposition  
\begin{equation}\label{eq:W-sum-match}
W_{ab} \;=\; \sum_{i=1}^m W^{(i)}_{ab},
\end{equation}  
in agreement with Eq.~(143) of \cite{Peng:2023yym}. Each term $W^{(i)}_{ab}$ further splits into symmetric and antisymmetric parts, as in Eq.~(145) of \cite{Peng:2023yym}:  
\begin{equation}\label{eq:W-split-sym-antisym}
W^{(i)}_{ab} \;=\; W^{(i)}_{(ab)} \;+\; W^{(i)}_{[ab]} \,.
\end{equation}  
These components can be written explicitly as  
\begin{equation}\label{eq:Peng-146-Latin}
\begin{aligned}
W^{(i)}_{(ab)}
&= \frac{1}{2}\bigg[
\sum_{k=1}^{i}(-1)^{k-1} g_{ah}\Big(
\nabla_{e_1}\cdots\nabla_{e_{k-1}}\,
Q^{e_1\cdots e_{i-1}(h\,|\,cdef\,|}\,
\nabla_{b)}\,
\nabla_{e_k}\cdots\nabla_{e_{i-1}}\,R_{cdef}
\Big) \\
&\hspace{2.6cm}
+\;\Big(\nabla_{e}\,\hat U^{(i)}_{(ab)}{}^{e}
-\nabla_{e}\,\hat U^{(i)\,e}{}_{(ab)}\Big)
+\;\Big(\nabla_{e}\,\tilde U^{(i)}_{(ab)}{}^{e}
-\nabla_{e}\,\tilde U^{(i)\,e}{}_{ab}\Big)
\bigg] ,
\end{aligned}
\end{equation}    
\begin{equation}\label{eq:Peng-147-Latin}
\begin{aligned}
W^{(i)}_{[ab]}
&= \frac{1}{2}\bigg[
\sum_{k=1}^{i}(-1)^{k-1}\Big(
g_{ah}\nabla_{e_1}\cdots\nabla_{e_{k-1}}\,
Q^{e_1\cdots e_{i-1}[\,h\,|\,cdef\,|}\,
\nabla_{b]}\,
\nabla_{e_k}\cdots\nabla_{e_{i-1}}\,R_{cdef}
\Big) \\
&\hspace{2.6cm}
+\;\nabla_{e}\,\hat U^{(i)}_{[\,a\,|}{}^{e}{}_{|\,b]}
+\;\nabla_{e}\,\tilde U^{(i)}_{[ab]}{}^{e}
\bigg] ,
\end{aligned}
\end{equation}  
where,
\begin{align*}
\hat{U}^{cab}_{(i)} 
&= 4 \sum_{k=1}^{i} (-1)^{k} 
\big( \nabla_{\lambda_1} \cdots \nabla_{\lambda_{k-1}} 
Q^{\lambda_1 \cdots \lambda_{k-1} a \lambda_{k+1} \cdots \lambda_i b n r s} \big) \,
\nabla_{\lambda_{k+1}} \cdots \nabla_{\lambda_i} R^{c}{}_{n r s}, \numberthis \label{eq:Peng-58-U}
\\[6pt]
\tilde{U}^{cab}_{(i)} 
&= \sum_{k=1}^{i-1} \sum_{j=k+1}^{i} (-1)^{k} 
\big( \nabla_{\lambda_1} \cdots \nabla_{\lambda_{k-1}} 
Q^{\lambda_1 \cdots \lambda_{k-1} a \lambda_{k+1} \cdots \lambda_{j-1} b \lambda_{j+1} \cdots \lambda_i \tau n r s} \big) 
\nabla_{\lambda_{k+1}} \cdots \nabla_{\lambda_{j-1}} 
\nabla^{c} \nabla_{\lambda_{j+1}} \cdots \nabla_{\lambda_i} R_{\tau n r s}. \numberthis \label{eq:Peng-58-U-tilde}
\end{align*}
\medskip

Collecting the volume-element and algebraic metric-dependence pieces together with the curvature-derived contribution \eqref{eq:IBP-result}, the variation of the Lagrangian density can be written
\begin{equation}\label{eq:deltaS-final}
\delta(\sqrt{-g}L)
=\sqrt{-g}\,E_{ab}\,\delta g^{ab}
+\sqrt{-g}\,\nabla_m\Theta^m(\delta g),
\end{equation}
with filed equation
\begin{align} \label{generaltheorfield}
    E_{ab}\equiv-\tfrac12 g_{ab} L
+P_{a}{}^{cde}R_{bcde}
-2\nabla^{c}\nabla^{d}P_{cabd}
+W_{ab} =0.
\end{align}
Each $\nabla_{(F_i)}R$ carries $(i+2)$ derivatives of the metric. Since $L$ depends on these up to $i=m$, the coefficients $Q^{E_i}$ may involve up to $(m+2)$ derivatives of $g$. Acting with up to $i$ additional derivatives to form $P$ gives terms with as many as $2m+2$ derivatives. Finally, the $\nabla\nabla P$ contribution to $E_{ab}$ adds two more, so the field equations $E_{ab}=0$ can contain up to $2(m+2)$ derivatives of $g$. For example: $R^2$ ($m=0$) yields fourth order, $(\nabla R)^2$ ($m=1$) yields sixth order, etc.

\subsection{Equation for the deformation in Lovelock theory of gravity}
The field equation in vacuum for the Lovelock theories of gravity is given by
\begin{equation}
    \mathcal{R}^a_{\phantom{a}b} \equiv P_b^{\phantom{b}ijk} R^a_{\phantom{a}ijk} = 0.
\end{equation}
Let's now compute:
\begin{align}
\mathcal{L}_\xi \mathcal{R}^a_{\phantom{a}b}
&= \mathcal{L}_\xi \left( P_b^{\phantom{b}ijk} R^a_{\phantom{a}ijk} \right) \\
&=  \left( \left(  \mathcal{L}_\xi P_b^{\phantom{b}ijk} \right)R^a_{\phantom{a}ijk} +  P_b^{\phantom{b}ijk}  \left(  \mathcal{L}_\xi R^a_{\phantom{a}ijk} \right)  \right).
\end{align}
The Lagrangian of this theory depends on $g^{ab}$ and $R^a_{\phantom{a}bcd}$,
but not on the covariant derivative of $R^a_{\phantom{a}bcd}$.
So,
\begin{align}
\mathcal{L}_\xi \mathcal{R}^a_{\phantom{a}b}
&= \left( \mathcal{L}_\xi R^a_{\phantom{a}mcd} \right) P_b^{\phantom{b}mcd}
+ R^a_{\phantom{a}mcd} \, \mathcal{L}_\xi \left( \frac{\partial L}{\partial R^b_{\phantom{b}mcd}} \right) \\
&= \left( \mathcal{L}_\xi R^a_{\phantom{a}mcd} \right) P_b^{\phantom{b}mcd}
+ R^a_{\phantom{a}mcd} \left( 
\frac{\partial}{\partial R^b_{\phantom{b}mcd}} \left( \frac{\partial L}{\partial g^{a'b'}} \right) \mathcal{L}_\xi g^{a'b'}
+ \frac{\partial}{\partial R^b_{\phantom{a}mcd}} \left( \frac{\partial L}{\partial R^{a'}_{\phantom{a'}m'c'd'}} \right) \mathcal{L}_\xi R^{a'}_{\phantom{a'}m'c'd'}
\right). \label{eq:liederiv}
\end{align}
Let us first evaluate the Lie derivative of the Riemann tensor.
\begin{align*}
    \mathcal{L}_\xi R^a_{\;\;mcd} &= \nabla_c (\mathcal{L}_\xi \Gamma^a_{dm}) - \nabla_d (\mathcal{L}_\xi \Gamma^a_{cm}) \\
    &= \frac{1}{2} \nabla_c \left[ g^{ai} \left( - \nabla_i \mathcal{L}_\xi g_{dm} + \nabla_d \mathcal{L}_\xi g_{mi} + \nabla_m \mathcal{L}_\xi g_{di} \right) \right] -  \frac{1}{2} \nabla_d \left[ g^{ai} \left( - \nabla_i \mathcal{L}_\xi g_{cm} + \nabla_c \mathcal{L}_\xi g_{mi} + \nabla_m \mathcal{L}_\xi g_{ci} \right) \right] \\
    &= \frac{1}{2} g^{ai} \Big[ - \nabla_c \nabla_i t_{dm} + \nabla_c \nabla_d t_{mi} + \nabla_c \nabla_m t_{di}
+ \nabla_d \nabla_i t_{cm} - \nabla_d \nabla_c t_{mi} - \nabla_d \nabla_m t_{ci} \Big] \\
&= \frac{1}{2} \Big[ - \nabla_c \nabla^a t_{dm} + R^e_{\;mdc} t^a_e - R^a_{\;edc} t^e_m + \nabla_c \nabla_m t^a_d + \nabla_d \nabla^a t_{cm} - \nabla_d \nabla_m t^a_c \Big]
\end{align*}
Contracting both sides with $P_b^{\;\;mcd}$, renaming the repeated indices 
$c \to d$ and $d \to c$ in the first term, and using the antisymmetry of the 
Riemann tensor, we obtain
\begin{align}
 P_b^{\;\;mcd} \, \mathcal{L}_\xi R^a_{\phantom{a}mcd}
&= \frac{P_b^{\;\;mcd} }{2} \Big[ 2 \nabla_d \nabla^a t_{cm} - 2 \nabla_d \nabla_m t^a_c
+ R^e_{\phantom{a}mdc} \, t^a_e - R^a_{\phantom{a}edm} \, t^e_m \Big]. \\
&= P_{bm}^{\;\; \; \;cd} \left( \nabla_d \nabla^a t_c^m - \nabla_d \nabla^m t^a_c \right)
+ \tfrac{1}{2}  P_{b}^{\;\;mcd} \left( R^e_{\phantom{e}mdc} \, t^a_e - R^a_{\phantom{a}edc} \, t^e_m \right). \label{eq:contra1}
\end{align}
Substituting the above expression, Eq.~\eqref{eq:contra1}, into Eq.~\eqref{eq:liederiv}, along with the similar simplifications for the second and third terms, we get
\begin{align}
    \mathcal{L}_\xi \mathcal{R}^a_{\phantom{a}b} &= P_{bm}^{\;\; \; \;cd} \left( \nabla_d \nabla^a t_c^m - \nabla_d \nabla^m t^a_c \right)
+ \tfrac{1}{2}  P_{b}^{\;\;mcd} \left( R^e_{\phantom{e}mdc} \, t^a_e - R^a_{\phantom{a}edc} \, t^e_m \right) + R^a_{\phantom{a}mcd } \bigg(   2 \frac{\partial P_b^{\phantom{a}mcd }}{\partial  R^{a'm'}_{\phantom{a'm'}c'd' }} \nabla_{d'}\nabla^{a'} t_{c'}^{m'} - \frac{\partial P_b^{\phantom{a}mcd }}{\partial  g^{a'b'}} t^{a'b'} \bigg)
\end{align}
Using the field equation of the Lovelock theories of gravity on the left-hand side, we get
\begin{align}
     P_{bm}^{\;\; \; \;cd} \left( \nabla_d \nabla^a t_c^m - \nabla_d \nabla^m t^a_c \right)
+ \tfrac{1}{2}  P_{b}^{\;\;mcd} \left( R^e_{\phantom{e}mdc} \, t^a_e - R^a_{\phantom{a}edc} \, t^e_m \right) + R^a_{\phantom{a}mcd } \bigg(   2 \frac{\partial P_b^{\phantom{a}mcd }}{\partial  R^{a'm'}_{\phantom{a'm'}c'd' }} \nabla_{d'}\nabla^{a'} t_{c'}^{m'} - \frac{\partial P_b^{\phantom{a}mcd }}{\partial  g^{a'b'}} t^{a'b'} \bigg) = 0
\end{align}

\subsection{Equation for the deformation in the general theory }

Taking the Lie derivative of the field equation in the general theory, discussed in Eq.\eqref{generaltheorfield}, we get
\begin{align}
\Lie E_{ab}
&= -\tfrac12 (\Lie g_{ab}) L - \tfrac12 g_{ab} \Lie L
+ (\Lie P_{a}{}^{cde}) R_{bcde} + P_{a}{}^{cde} \Lie R_{bcde} -2 \Lie\big( \nabla^{c}\nabla^{d} P_{cabd} \big)
+ \Lie W_{ab} = 0.
\label{eq:LxiE-master}
\end{align}
Now we compute the Lie derivative of different terms in Eq.\eqref{eq:LxiE-master}. Defining $t_{ab} \equiv \Lie g_{ab} = 2 \nabla_{(a} \xi_{b)}$ and using
\begin{align}
\Lie g^{ab} & = - t^{ab},
\qquad
\Lie \sqrt{-g} = \tfrac12 \sqrt{-g}\, g^{ab} t_{ab},
\\
\Lie L& = \frac{\partial L}{\partial g^{pq}} \Lie g^{pq}
+ \sum_{i=0}^{m} Q^{F_i cdef} \Lie(\nabla_{(F_i)} R_{cdef})
= - \frac{\partial L}{\partial g^{pq}} t^{pq}
+ \sum_{i=0}^{m} Q^{F_i cdef} \Lie(\nabla_{(F_i)} R_{cdef}) ,
\end{align}
we obtain
\begin{align}
\Lie E_{ab}
&= -\tfrac12 t_{ab} L
+ \tfrac12 g_{ab} \frac{\partial L}{\partial g^{pq}} t^{pq}
- \tfrac12 g_{ab} \sum_{i=0}^m Q^{F_i cdef} \Lie(\nabla_{(F_i)} R_{cdef})
\label{eq:LxiE-expanded}\\
&\quad + (\Lie P_{a}{}^{cde}) R_{bcde} + P_{a}{}^{cde} \Lie R_{bcde}
-2 \Lie(\nabla^{c}\nabla^{d} P_{cabd})
+ \Lie W_{ab} =0.
\nonumber
\end{align}
Starting from the definition of the Lie derivative acting on the covariant derivative of a \((p,q)\)-tensor \(T\):
\begin{align*}
\mathcal{L}_\xi \!\big( \nabla_f T^{a_1 \cdots a_p}{}_{b_1 \cdots b_q} \big)
&= \xi^c \nabla_c \!\left( \nabla_f T^{a_1 \cdots a_p}{}_{b_1 \cdots b_q} \right) \quad - (\nabla_c \xi^{a_1}) \, \nabla_f T^{c a_2 \cdots a_p}{}_{b_1 \cdots b_q}
- \cdots  + (\nabla_{b_1} \xi^c) \, \nabla_f T^{a_1 \cdots a_p}{}_{c b_2 \cdots b_q}
+ \cdots, \numberthis
\end{align*}
subtracting  \(\nabla_f (\mathcal{L}_\xi T^{a_1 \cdots a_p}{}_{b_1 \cdots b_q})\) from the left-hand side, and using identity
\begin{align}
\mathcal{L}_\xi \Gamma^{a}{}_{bc}
= \nabla_b \nabla_c \xi^a
  + R^{a}{}_{c b d}\, \xi^d \, 
\end{align}
we obtain \cite{yano1957lie, unknown, Lompay:2013opa}: 
\begin{align*}
\mathcal{L}_\xi \!\left( \nabla_f 
  T^{a_1 \cdots a_p}{}_{b_1 \cdots b_q} \right)
= \nabla_f \!\left( 
  \mathcal{L}_\xi T^{a_1 \cdots a_p}{}_{b_1 \cdots b_q} \right)
+ \sum_{i=1}^p 
  \left( \mathcal{L}_\xi \Gamma^{a_i}{}_{f e} \right) 
  T^{a_1 \cdots e \cdots a_p}{}_{b_1 \cdots b_q}
- \sum_{j=1}^q 
  \left( \mathcal{L}_\xi \Gamma^{e}{}_{f b_j} \right) 
  T^{a_1 \cdots a_p}{}_{b_1 \cdots e \cdots b_q} \, . \numberthis \label{generalident}
\end{align*}
Using the above general identity we get 
\begin{align*}
\mathcal{L}_\xi \big( \nabla_f R_{amcd} \big) 
&= \nabla_f \big( \mathcal{L}_\xi R_{amcd} \big) 
+ \big( \mathcal{L}_\xi \Gamma^a_{\,\,fe} \big) R^{e}{}_{mcd}
- \big( \mathcal{L}_\xi \Gamma^e_{\,\,fm} \big) R_{aecd}
- \big( \mathcal{L}_\xi \Gamma^e_{\,\,fc} \big) R_{amed}
- \big( \mathcal{L}_\xi \Gamma^e_{\,\,fd} \big) R_{amce}. \\
&= \nabla_f(\Lie R^{a}{}_{mcd})
+ \tfrac12 g^{as}(\nabla_f t_{se} + \nabla_s t_{fe} - \nabla_e t_{fs}) R^{e}{}_{mcd} - \tfrac12 g^{es}(\nabla_f t_{ms} + \nabla_m t_{fs} - \nabla_s t_{fm}) R^{a}{}_{ecd} \nonumber\\
&\quad - \tfrac12 g^{es}(\nabla_f t_{cs} + \nabla_c t_{fs} - \nabla_s t_{fc}) R^{a}{}_{med} - \tfrac12 g^{es}(\nabla_f t_{ds} + \nabla_d t_{fs} - \nabla_s t_{fd}) R^{a}{}_{mce} \numberthis,
\end{align*}
where
\begin{align}
\Lie R_{abcd}
&= \tfrac12 \Big(
 \nabla_c \nabla_b t_{ad}
+\nabla_d \nabla_a t_{bc}
-\nabla_c \nabla_a t_{bd}
-\nabla_d \nabla_b t_{ac}
\Big)
+ t_{e[a} R^{e}{}_{b]cd} + t_{e[c} R^{e}{}_{d]ab}.
\label{eq:LxiRabcd-basic}
\end{align}
Repeating the same procedure for higher symmetrized derivatives one gets:
\begin{align} \label{eq:LxiHigherDR}
\Lie\big(\nabla_{(e_1}\cdots\nabla_{e_i)} R_{abcd}\big)
&= \nabla_{(e_1}\cdots\nabla_{e_i)} (\Lie R_{abcd})
+ \sum_{r=1}^{i} \mathcal{C}^{(r)}_{e_1\cdots e_i\,abcd}[ \nabla^{(\le r)} t, \nabla^{(\le i-r)} R ], 
\end{align}
with each $\mathcal{C}^{(r)}$ containing at most $r$ derivatives on $t$.
Using chain rule we get
\begin{align}
\Lie P^{a}{}_{cde}
= \frac{\partial P^{a}{}_{cde}}{\partial g^{pq}} (-t^{pq})
+ \sum_{j=0}^m
  \frac{\partial P^{a}{}_{cde}}{\partial(\nabla_{(F_j)} R_{ghij})}
  \Lie(\nabla_{(F_j)} R_{ghij}).
\end{align}
For the double divergence:
\begin{align}
-2\,\Lie\!\big(\nabla^{c}\nabla^{d} P_{ c a b d}\big)
&= -2\,\nabla^{c}\nabla^{d}\big(\Lie P_{ c a b d}\big)
+ 2\,t^{c}{}_{e}\,\nabla^{e}\nabla^{d} P_{ c a b d}
+ 2\,t^{d}{}_{f}\,\nabla^{c}\nabla^{f} P_{ c a b d}
+ \mathcal{J}_{ab}\big[P;\nabla t,\nabla\nabla t\big],
\end{align}
with $\mathcal{J}_{ab}$ given by
\begin{align}
\mathcal{J}_{ab}[P;\nabla t,\nabla\nabla t]
&\equiv -2\Big\{
(\Lie\Gamma)^{e}{}_{c}{}^{c}\,\nabla^{d} P_{e a b d}
+ 2\,(\Lie\Gamma)^{d}{}_{c e}\,\nabla^{c} P^{e}{}_{a b d}
+ \big(\nabla^{c}(\Lie\Gamma)^{d}{}_{c e}\big)\, P^{e}{}_{a b d}
\Big\}.
\label{eq:Jab-expanded}
\end{align}

with
\begin{align}
\mathcal{L}_\xi \Gamma^p_{\,\,qr} 
= \tfrac{1}{2} g^{ps} \big( \nabla_q t_{rs} + \nabla_r t_{qs} - \nabla_s t_{qr} \big).
\end{align}

Since both $W^{(i)}_{(ab)}$ and $W^{(i)}_{[ab]}$ are constructed covariantly from
the metric and from covariant derivatives of the curvature (entering partly through
the symmetrised blocks $\nabla_{(F_j)}R$ that define the $Q$'s, and partly through
unsymmetrised derivative strings in the $\hat U^{(i)}$ and $\tilde U^{(i)}$ terms),
the Lie derivative of $W_{ab}$ can be written as
\begin{equation}\label{eq:LieW-chain-corrected}
\mathcal{L}_\xi W_{ab}
=
\frac{\partial W_{ab}}{\partial g^{pq}}\big(-t^{pq}\big)
+ \sum_{\mathcal{I}}
\frac{\partial W_{ab}}{\partial T_{\mathcal{I}}}\;
\mathcal{L}_\xi T_{\mathcal{I}},
\end{equation}
where the set $\{T_{\mathcal{I}}\}$ runs over the
independent curvature--derivative tensors that actually appear in $W_{ab}$. We note that each term in the set $\mathcal{L}_\xi T_{\mathcal{I}}$ depends on $t_{ab}$ linearly and are $0$ when $t_{ab} = 0$. To estimate the derivative order, note that $R_{abcd}$ involves up to two derivatives
of the metric, so $\nabla_{(F_j)}R_{abcd}$ contains up to $j+2$ derivatives of $g_{ab}$.
Under a metric variation $\delta g_{ab}=\mathcal{L}_\xi g_{ab}=t_{ab}$, the linearized
variation of the curvature contains two derivatives of $\delta g$, i.e.\ $\delta R\sim\nabla\nabla t$. Therefore varying any (symmetrised or unsymmetrised) derivative string of $R$ yields terms with up to $j+2$ covariant derivatives acting on $t_{ab}$. Since $j\le m$, it follows that $\mathcal{L}_\xi W_{ab}$ is built from $t_{ab}$ and its covariant derivatives, with at most $m+2$ derivatives of $t_{ab}$ in the worst case (equivalently, up to $m+3$ derivatives of $\xi^a$).

Gathering terms by their structure in $t_{ab}$ we can reorganize $\mathcal{L}_\xi E_{ab} =0$, in Eq.(\ref{eq:LxiE-expanded}) and get

\begin{align}
&P_{a}{}^{mcd}\Big(\nabla_{d}\nabla_{b}t_{mc}-\nabla_{d}\nabla_{m}t_{bc}\Big)
+\tfrac12\,P_{a}{}^{mcd}\Big(R^{e}{}_{mdc}\,t_{be}-R_{b\;dc}^{\;e}\,t_{me}\Big)
\nonumber\\[2pt]
&\quad
+ R_{b m c d}\Big(
  2\,\frac{\partial P_{a}{}^{mcd}}{\partial R_{a' m'c'd'}}\;\nabla_{d'}\nabla_{a'} t_{m'c'}
  - \frac{\partial P_{a}{}^{mcd}}{\partial g^{a'b'}}\,t^{a'b'}
\Big)
-\tfrac12\, t_{ab}\,L
+ \tfrac12\, g_{ab}\,\frac{\partial L}{\partial g^{a'b'}}\,t^{a'b'}
\nonumber\\[2pt]
&\quad
-g_{ab}P^{cdef}\bigg(\nabla_e \nabla_d t_{cf} + \frac{1}{2}\bigg(t_{e'[c}R^{e'}_{\;\;d]ef} + t_{e'[e}R^{e'}_{\;\;f]cd}\bigg)\bigg)
\nonumber\\[2pt]
&\quad
-2\Bigg[
 2\,\nabla^{m}\nabla^{n}\!\left(\frac{\partial P_{manb}}{\partial R_{a' m'c'd'}}\right)\,\nabla_{d'}\nabla_{a'} t_{m'c'}
 - \nabla^{m}\nabla^{n}\!\left(\frac{\partial P_{manb}}{\partial g^{a'b'}}\right)\,t^{a'b'}
 + \cdots
\Bigg]
\nonumber\\[2pt]
&\quad
+ 2\,t^{c}{}_{e}\,\nabla^{e}\nabla^{d} P_{c a b d}
+ 2\,t^{d}{}_{f}\,\nabla^{c}\nabla^{f} P_{ c a b d}
+ \,\mathcal{J}_{ab}\big[P;\nabla t,\nabla\nabla t\big]
+ \mathcal{W}_{ab}\big[t,\nabla t,\ldots,\nabla^{m+2}t\big] =0.
\label{eq:FinalDeformationEqcorrected}
\end{align}
Here, $\mathcal{W}_{ab}$ collects all terms descending from $\Lie W_{ab}$ and from $\Lie(\nabla_{(E_i)} R)$ with $i\ge 1$. The ellipsis “$\cdots$” encompasses further contributions arising from differentiating $P$ with respect to higher symmetrised curvature derivatives. The above equation \eqref{eq:FinalDeformationEqcorrected} is linear in $t_{ab}$. When interpreting $\mathcal{L}_\xi E_{ab}$ as a differential operator on the deformation tensor $t_{ab}=\mathcal{L}_\xi g_{ab}$, all contributions have a common structure:
each $\mathcal{L}_\xi(\nabla_{(F_j)}R)$ produces at most $(j+2)$ derivatives of $t_{ab}$,
since $\mathcal{L}_\xi R\sim\nabla\nabla t + tR$ carries two derivatives of $t$ and the
$j$ symmetrised covariant derivatives add at most $j$ more. The maximal case $j=m$ gives
$(m+2)$ derivatives of $t$. Any extra covariant derivatives outside such as the term $\nabla^c\nabla^d$ in 
\begin{align}
-2\nabla^c\nabla^d(\tfrac{\partial P}{\partial(\nabla_{(F_m)}R)}
\,\mathcal{L}_\xi(\nabla_{(E_m)}R)),
\end{align}
or the analogous derivatives in $\mathcal{W}_{ab}$ and related terms, can increase the order by at most two. Therefore, regardless of which
piece of $E_{ab}$ one considers, the Lie derivative $\mathcal{L}_\xi E_{ab}$ contains at
most $(m+4)$ derivatives of $t_{ab}$. In the generic situation—absent Lovelock-type
cancellations or Bianchi-type identities—this bound is saturated, so the principal part
of $\mathcal{L}_\xi E_{ab}$ has differential order $m+4$ acting on $t_{ab}$.

We notice that $t_{ab}=0$ is a solution of Eq.\eqref{eq:FinalDeformationEqcorrected}. To show its uniqueness, with vanishing Cauchy data at the stellar surface, in the next subsection  we follow an approach similar to the case of Lovelock theories discussed in the main text.

\subsection{Uniqueness for the general theory}
We consider a smooth hypersurface $\Sigma$ that separates the stellar interior $\mathcal{M}_1$ from the exterior region $\mathcal{M}_2$. 
Let $\xi^a$ denotes the Killing field in $\mathcal M_1$ tangent to $\Sigma$, and we extend it near $\Sigma$ by the vector wave equation
\begin{equation}
\nabla_a \nabla^a \xi^b =- R^b{}_c \,\xi^c .
\label{wave}
\end{equation}
Writing Eq.\eqref{wave} in coordinates, we have
\begin{equation}
g^{\mu\nu}\partial_\mu \partial_\nu \xi^b 
= - B^b(\xi,\partial \xi,g,\partial g) - \,\xi^{d}\partial_{d}\!\left(g^{ce}\Gamma^{b}{}_{ce}\right) ,
\label{coord}
\end{equation}
where $B^b$ depends on $\partial g$ (via $\Gamma$) and $\partial^2 g$ (via $\partial \Gamma$). Since the RHS of \eqref{coord} is continuous under our assumptions, 
it follows that $\partial^2 \xi$ and thus $\nabla t$ are continuous.

Differentiating \eqref{wave} and commuting derivatives yields schematically
\begin{align}
\nabla^3 \xi = (\nabla R)\cdot \xi + R \cdot \nabla \xi .
\end{align}
We also assume that the hypersurface $\Sigma$ is non-characteristic for the wave operator $\Box_g$, so that the principal part $g^{\mu\nu}\partial_\mu \partial_\nu$ can be inverted to solve for the highest normal derivative. After $(r-1)$ differentiations, the coordinate version reads
\begin{align}
g^{\mu\nu}\partial_\mu \partial_\nu (\partial^{\,r-1}\xi)
= F_r\Big(\xi,\partial \xi,\ldots,\partial^r \xi;\, 
       g,\partial g,\ldots,\partial^{r+1} g;\,
       R,\nabla R,\ldots,\nabla^{r-1}R \Big),
\end{align}
where the derivative-counting is consistent because: differentiating Christoffel symbols up to $(r-1)$ times produces up to $\partial^{r+1} g$, the curvature derivatives $\nabla^{\,r-1} R$ also involve $\partial^{r+1} g$.
Now, if $g \in C^{m+3}$, and no components with differential order $m+4$ that appear are discontinuous, similar to the case in Lovelock theories, then $\nabla^{r+1}\xi$, and hence $\nabla^{r} t$, are continuous across $\Sigma$ for all $r \le m+3$. Since $t=0$ in $\mathcal M_1$, continuity gives
\begin{align}
\nabla^r t_{ab}\big|_{\Sigma}=0 \qquad (\forall r\le m+3).
\end{align}

Equivalently, one may view $\mathcal{L}_\xi E_{ab}$ as a linear differential operator in $t_{ab}=\mathcal{L}_\xi g_{ab}$ by replacing $\mathcal{L}_\xi g\mapsto t$ and $\mathcal{L}_\xi\Gamma\mapsto \tfrac12\,g^{-1}\!\ast\nabla t$, yielding a homogeneous equation $\mathsf{G}_{ab}[t]=0$ with principal order $m{+}4$. On the spacelike, non-characteristic hypersurface $\Sigma$, then the normal derivatives vanish up to order $(m{+}3)$,
\begin{align}
(n\!\cdot\!\nabla)^k t_{ab}\big|_{\Sigma}=0 \qquad (k=0,1,\ldots,m{+}3),
\end{align}
and the intrinsic constraints induced on $\Sigma$ by $\mathsf{G}_{ab}[t]=0$ hold. This bootstrap provides the full vanishing jet $\nabla^r t|_\Sigma=0$ for all $r\le m{+}3$, ensuring the sufficient initial data for the Cauchy problem of order-$(m{+}4)$ $t$–equations. The uniqueness theorem of Holmgren again yields $t_{ab}\equiv0$ just outside the surface of the star, as $t=0$ is a solution of the system of differential equations shown in Eq.\eqref{eq:FinalDeformationEqcorrected}.

\bibliography{apssamp}